\documentclass[iop]{emulateapj}

\def\lapp{\ifmmode\stackrel{<}{_{\sim}}\else$\stackrel{<}{_{\sim}}$\fi}
\def\gapp{\ifmmode\stackrel{>}{_{\sim}}\else$\stackrel{>}{_{\sim}}$\fi}

\begin{document}

\submitted{Accepted June 27, 2011 }

\title{The 2009 outburst of magnetar 1E 1547$-$5408: Persistent radiative and burst properties}

\author{
P. Scholz\altaffilmark{1}
\& V. M. Kaspi\altaffilmark{1}
}

\altaffiltext{1}{Department of Physics, Rutherford Physics Building,
McGill University, 3600 University Street, Montreal, Quebec,
H3A 2T8, Canada}

\begin{abstract}

The magnetar 1E~1547$-$5408 recently exhibited two periods of outburst, beginning on 2008 October 3 and 2009 January 22.
Here we present an analysis of the persistent radiative evolution and a statistical study of the burst properties during the 2009 outburst
using the {\em Swift} X-ray Telescope (XRT). We find that the 1--10 keV flux increased by a factor of $\sim500$
and hardened significantly, peaking $\sim6$ hours after the onset of the outburst. 
The observed pulsed fraction exhibited an anti-correlation with phase-averaged flux. Properties of the several hundred X-ray bursts during the 2009 outburst were determined and compared to those 
from other magnetar outburst events. 
We find that the peaks of the bursts occur randomly in phase but that the folded counts that compose 
the bursts exhibit a pulse which is misaligned with the persistent pulse phase. 
We also report a correlation between burst hardness and flux.
We compare the hardness-flux evolution 
of the persistent emission of both outbursts
to those from other magnetars and find that although there does exist an overall trend, the degree of hardening for a given 
increase in flux is not uniform from source to source. These results are discussed in the context of previous results and 
within the magnetar model.

\end{abstract}

\keywords{stars: neutron --- pulsars: individual (1E 1547$-$5408, PSR J1550$-$5418, SGR J1550$-$5418) --- X-rays: bursts --- X-rays: general}

\section{Introduction}

Originally classified as two distinct types of objects, anomalous X-ray pulsars (AXPs) and soft gamma repeaters (SGRs) are now generally 
accepted to be highly magnetized neutron stars, magnetars, with magnetic fields $B > 10^{14} - 10^{15}$ G \citep[for reviews, see][]{wt06,mer08}. 
Magnetars differ from `normal' rotation-powered pulsars in that their X-ray luminosities exceed the rate of energy released from their spin-down. 
They are believed to be powered by the decay of their magnetic fields. Another common property of magnetars is that they exhibit episodes 
of violent bursting activity. At one point thought to be only observed in SGRs, X-ray bursts in AXPs were first detected from 1E 1048.1$-$5937 in 
late 2001 \citep{gkw02}. Bursts have since been detected in several other AXPs \citep{kgw+03,wkg+05,kbc+06,gdk09}.


1E 1547$-$5408 was first discovered as an X-ray source by the {\em Einstein} satellite \citep{lm81}. It was identified only recently as a magnetar by
\citet{gg07} based on its X-ray spectrum and infrared flux, as well as its possible association with the supernova remnant G327.24$-$0.13. Pulsed radio emission was detected by
\citet{crhr07} at a period of $\sim$2 s, the shortest period of all known magnetars\footnote
{see the McGill SGR/AXP Online Catalog, http://www.physics.mcgill.ca/$\sim$pulsar/magnetar/main.html}. 
An {\em XMM-Newton} observation in 2007 showed a significant flux enhancement from what was previously measured \citep{hgr+08}, thus revealing that an X-ray outburst
event had occured between 2006 and 2007. Another outburst event occured on 2008 October 3 and was detected by the {\em Swift} \citep{ier+10} and {\em Fermi} 
\citep{kgk+10} satellites. 1E 1547$-$5408 entered yet another active phase on 2009 January 22 when hundreds of bursts were detected by {\em Swift}, {\em INTEGRAL}, 
and {\em Fermi}. \citet{tve+10} report the appearance of dust scattering rings around 1E 1547$-$5408 after the 2009 burst in follow-up observations with {\em XMM-Newton} and {\em Swift}. 
The energy released by the event causing the rings was estimated to be $10^{44} - 10^{45}$ ergs. 
From these rings, a distance to the source was determined to be 3.9 kpc. 
\citet{nkd+10} report on {\em Chandra} observations beginning $\sim 2$ days following the outburst. They note a lack of spectral variation during the flux decay as 
well as an anti-correlation between pulsed fraction and phase-averaged flux.

The origin and flux evolution of magnetar outbursts are not presently well understood \citep[but see][]{pp11}, and
the sample of observed outbursts is still relatively small \citep[for a review, see][]{re11}.  Hence, it is important for each
magnetar outburst to be promptly observed and studied.  The magnetar model suggests that outbursts
are a result of magnetospheric twists of the magnetic field structure following some form
of internal energy or stress release \citep{tlk02,bel09}, which predicts
a correlation between hardness and X-ray flux in magnetar outbursts.
Indeed a universal relationship between flux increase and spectral hardening might
be expected, if not from source to source, at least for individual sources.  Also, the origin and
physics of X-ray bursts from magnetars are poorly understood.  Bursts have been proposed to be
magnetospheric in origin \citep{lyu03} as well as originating from stresses in the crust of the neutron star \citep{td95}.  
Detailed statistical studies of magnetar bursts can help determine their
properties for comparison with model predictions, yet have only been done for three sources
thus far \citep{gkw+01,gkw04}.


This paper presents the results of an analysis of the persistent emission from the 2009 event as well as a statistical study of the bursts using observations 
from {\em Swift}. The burst study will
focus primarily on bursts from the 2009 outburst since the number ($\sim$ 400) is much higher than in the 2008 event, in which 8 bursts were detected by
{\em Swift} \citep{ier+10}. A summary of the observations is presented in \S2. The analysis performed on the data as well as the results for the persistent emission 
are reported in \S3.1. The analysis and results for the burst study are presented in \S3.2. The results and their possible physical interpretations are discussed in \S4. 
Finally, our findings are summarized in \S5.

\section{Observations}

The data presented in this paper were obtained using the X-Ray Telescope (XRT) \citep{bhn+05} on the {\em Swift} satellite.
The XRT uses Wolter-I optics and an {\em XMM-Newton}/EPIC MOS CCD detector to provide rapid imaging and spectra of X-ray transients in the 0.5--10 keV energy range.
During the 2008 outburst of 1E 1547$-$5408, the {\em Swift} Burst Alert Telescope (BAT) triggered at 09:28:08 UT on 2008 October 3 and {\em Swift} promptly slewed to
the source position. The XRT began taking observations 99 s after the trigger. On 2009 January 22, the BAT triggered
at 01:32:41 UT and the first XRT observation began $\sim$50 min later. Table \ref{ta:obs09} shows a
summary of the observations following the 2009 outburst event as well as two observations preceding the outburst.
The observations following the 2008 outburst were previously presented by \citet{ier+10}; here we focus on the 2009 event.

Cleaned data products in both windowed-timing (WT) and photon-counting (PC) modes were obtained from the HEASARC {\em Swift} Archive. 
Data were reduced to the barycentre using the
position of 1E 1547$-$5408, 15$\mathrm{^h}$50$\mathrm{^m}$54\fs11, $-$54\degr18\arcmin23\farcs 7 \citep{crhr07}. 
For WT mode, a source region consisting of a 40-pixel-long strip centered on the pulsar position was extracted. 
A background region of the same size was extracted on a source-free position away from 1E 1547$-$5408. For the PC mode data, an annulus 
with outer radius 20 pixels and 4 pixel inner radius was extracted for the source, and an
annular region of 20 pixel inner radius and 50 pixel outer radius was used for the background. The 4 pixel inner region was excluded 
to avoid pileup.

\section{Analysis and Results}

\subsection{Persistent flux evolution}

In order to investigate the behavior of the persistent flux of 1E 1547$-$5408 during the 2009 outburst,
the observations were fitted with spectral models. Spectra were extracted from the event lists using {\ttfamily xselect}.
Spectral fitting was performed using the XSPEC\footnote{http://xspec.gfsc.nasa.gov}  package version 12.6. 
The spectra were grouped with a minimum of 20 counts per energy bin. Ancillary response files were created
using the FTOOL {\ttfamily xrtmkarf} and the standard spectral redistribution matrices from the {\em Swift} CALDB were used.
Bursts were then removed from the observations using the method described in \S\ref{sec:bursts}. 

The first two observations following the BAT trigger were 
split into shorter intervals $\sim$2~ks in length, since the spectral properties were evolving rapidly during that time.
These split observations occured during the first day of the outburst when the dust scattering rings \citep{tve+10} were not fully resolved by the {\em Swift} XRT. 
For these observations, the radiative properties of the source cannot be simply disentangled from the delayed emission from the dust scattering rings 
(A. Tiengo \& P. Esposito, private communication). 
Appropriate modelling of this is currently under investigation (Tiengo et al., in preparation) but beyond the scope of our work.
Following the first day after the outburst, the dust scattering rings were outside of the extraction region and so those observations 
are not affected by dust scattering.

\begin{figure}
\plotone{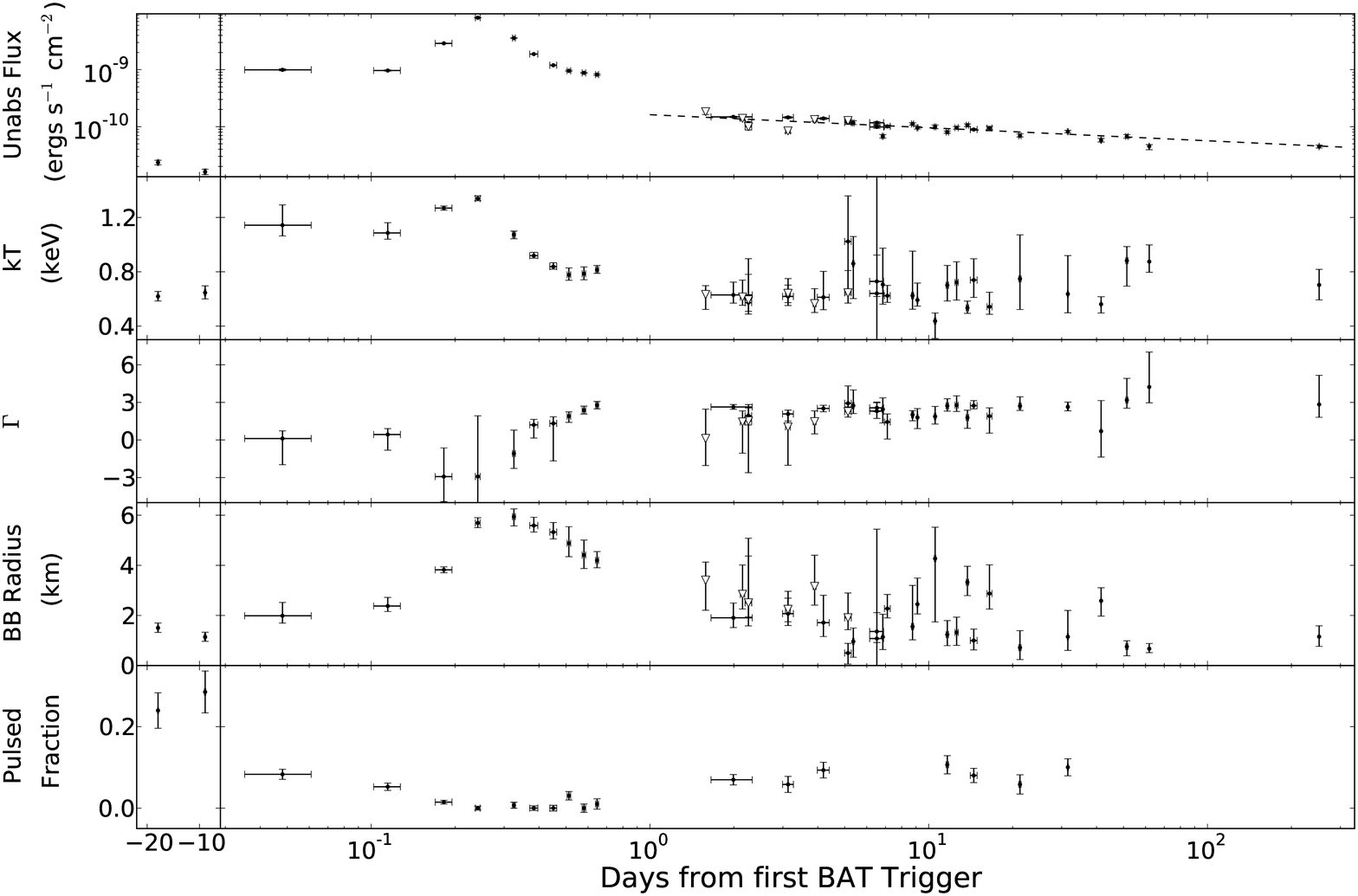}
\figcaption{Properties of the persistent emission of 1E 1547$-$5408 surrounding the 2009 outburst event.
The panels on the left show the two observations that preceded the event, on a linear time scale. The right-side
panels are the post-event observations and are plotted on a logarithmic time scale. 
The dashed line in the top panel shows the power-law decay fit to the unabsorbed flux.
The unabsorbed flux in the top panel
and the pulsed fraction in the bottom panel are for the 1--10 keV range. 
The first ten points (the first day) after the BAT trigger are contaminated by dust scattering and should 
thus be regarded with caution.
Some of the early PC mode observations had background regions that were contaminated by the dust scattering;
these are marked with an open triangle.
\label{fig:spec}
}
\end{figure}

The spectra were fitted with a photoelectrically absorbed blackbody with an added 
power-law component. To determine $N_H$, we fit observations having an exposure time
greater than 3 ks jointly with a single $N_H$. The parameters $kT$, $\Gamma$ and their normalizations were allowed to vary in these fits. 
This resulted in $N_H = 3.24(5) \times 10^{22}$ cm$^{-2}$ which is consistent with the value of $3.1_{-0.8}^{+0.7} \times 10^{22}$ cm$^{-2}$ reported in \citet{gg07}, 
though is somewhat lower than the value of $4.1(1) \times 10^{22}$ cm$^{-2}$ measured by \citet{nkd+10}.
All the {\em Swift} observations were subsequently fit with the column density fixed to our best-fit value.
The 2008 data have been previously presented by \citet{ier+10}, whose results are generally consistent with those of our analysis for the same time period,
so they will not be presented here. Figure \ref{fig:spec} shows the results of fitting of the observations 
following the 2009 outburst. 
The split observations, affected by dust scattering, are the first 10 data points in Figure \ref{fig:spec} following the trigger,
and so should be regarded with caution. 
The two observations preceding the outburst were fit with only a blackbody and no
power-law component. This is because the power-law index could not be constrained due to a paucity of counts.
The fits in general were excellent; the goodness-of-fit statistic $\chi^2_\nu$ ranged between 0.61 and 1.48 with a mean of 1.06. 


The peak of the persistent emission occured $\sim$6 hr after the BAT first triggered on 1E 1547$-$5408 in 2009 January. The
peak was almost 3 orders of magnitude higher in flux than the pre-outburst emission.
This was accompanied by a hardening of the spectrum in the 1-10 keV band, as can be seen in the falling spectral index and rising $kT$.
After the peak, the spectral index softened and $kT$ fell as the flux dropped. 
To characterize the flux decay of the outburst, a power law decay 
was fit to the data following the first day of the outburst. 
The first day was not included as the source decay is superimposed by delayed emission from the dust scattering rings.
The power law decay is descibed by $F = A(t-t_0)^{\alpha}$, where $F$ is the unabsorbed flux, 
 $A$ is the normalization, $\alpha$ is the power-law index, and $t_0$ is the time of the BAT trigger. 
The decay was described by a power-law with an index of $-0.24 \pm 0.02$.
Although the $\chi_\nu^2/\nu$ of the fit was 4.1/23, a power-law model fit the data much better than an exponential decay.
To determine the blackbody radius of the emitting region as shown in Figure \ref{fig:spec}, a distance of $3.91\pm0.07$ kpc from \citet{tve+10} was assumed.


To measure pulsed fractions and fluxes, the burst-removed time series (see \S3.2) were folded at the rotational ephemeris derived from 
contemporaneous {\em RXTE} observations of 1E 1547$-$5408 \citep{dkg11}. Specifically,
we used spin frequency $\nu = 0.48259615(3)$ Hz, with frequency derivative 
$\dot{\nu} = -5.12(2) \times 10^{-12}$ s$^{-2}$ at reference epoch MJD 54854.0 or 2009 January 23. 
The pulsed flux was calculated using an RMS method according to the formula in \citet{dkg08}, with 7 harmonics.
The pulsed fraction was determined by dividing the pulsed flux by the phase-averaged flux.
A pulsed flux and fraction were measured for each WT observation with an exposure time $>3$ ks for which the ephemeris was valid.
As seen in Figure \ref{fig:spec}, the pulsed fraction decreased as the flux increased.
After $\sim$40 days, the pulsed fraction had not yet recovered to its pre-burst level of $\sim$0.25.

\begin{figure}
\plotone{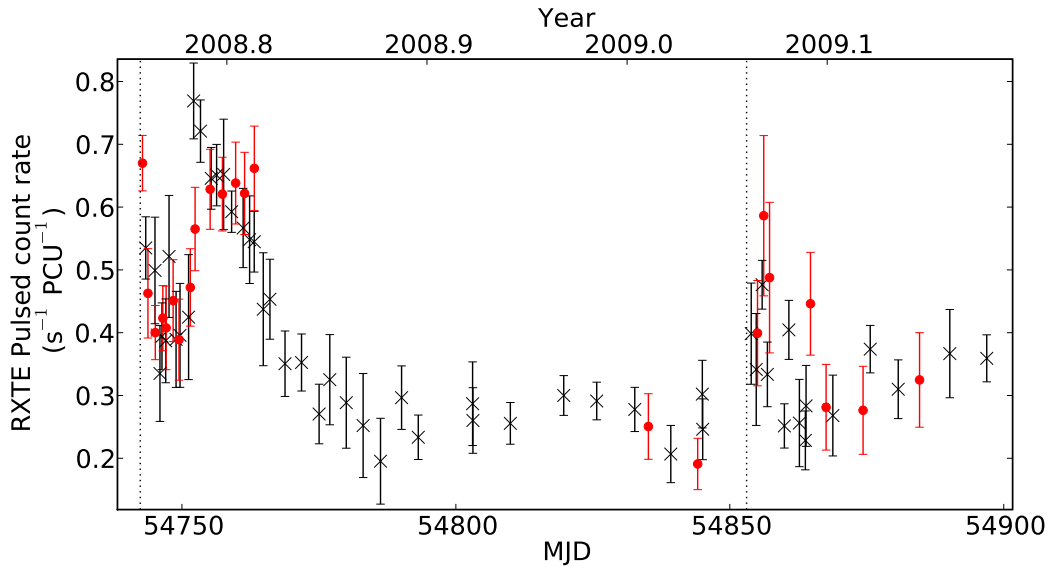}
\figcaption{2-10 keV RMS pulsed flux evolution of 1E 1547$-$5408 determined from {\em Swift} XRT and {\em RXTE}.
The black crosses show the {\em RXTE} pulsed count rates and the red points are the {\em Swift} pulsed count rates,
arbitrarily scaled to the {\em RXTE} values. The dotted vertical lines mark the onsets of the 2008 and 2009 outbursts.
\label{fig:pflux}
}
\end{figure}

The pulsed flux evolution around both the 2008 and 2009 outburst events is presented in Figure \ref{fig:pflux}.
The {\em Swift} data confirm the evolution that is observed in the {\em RXTE} data which are presented briefly in \citet{nkd+10}
and in detail in \citet{dkg11}.
The pulsed flux enhancement $\sim$11 days following the initial trigger of the 2008 event, although not noted by
\citet{ier+10}, is clearly present in the {\em Swift} data. 
Thus, the 2009 event showed a much smaller increase in pulsed flux than did the 2008 event,
while the opposite is true of the total flux.

\begin{figure}
\plotone{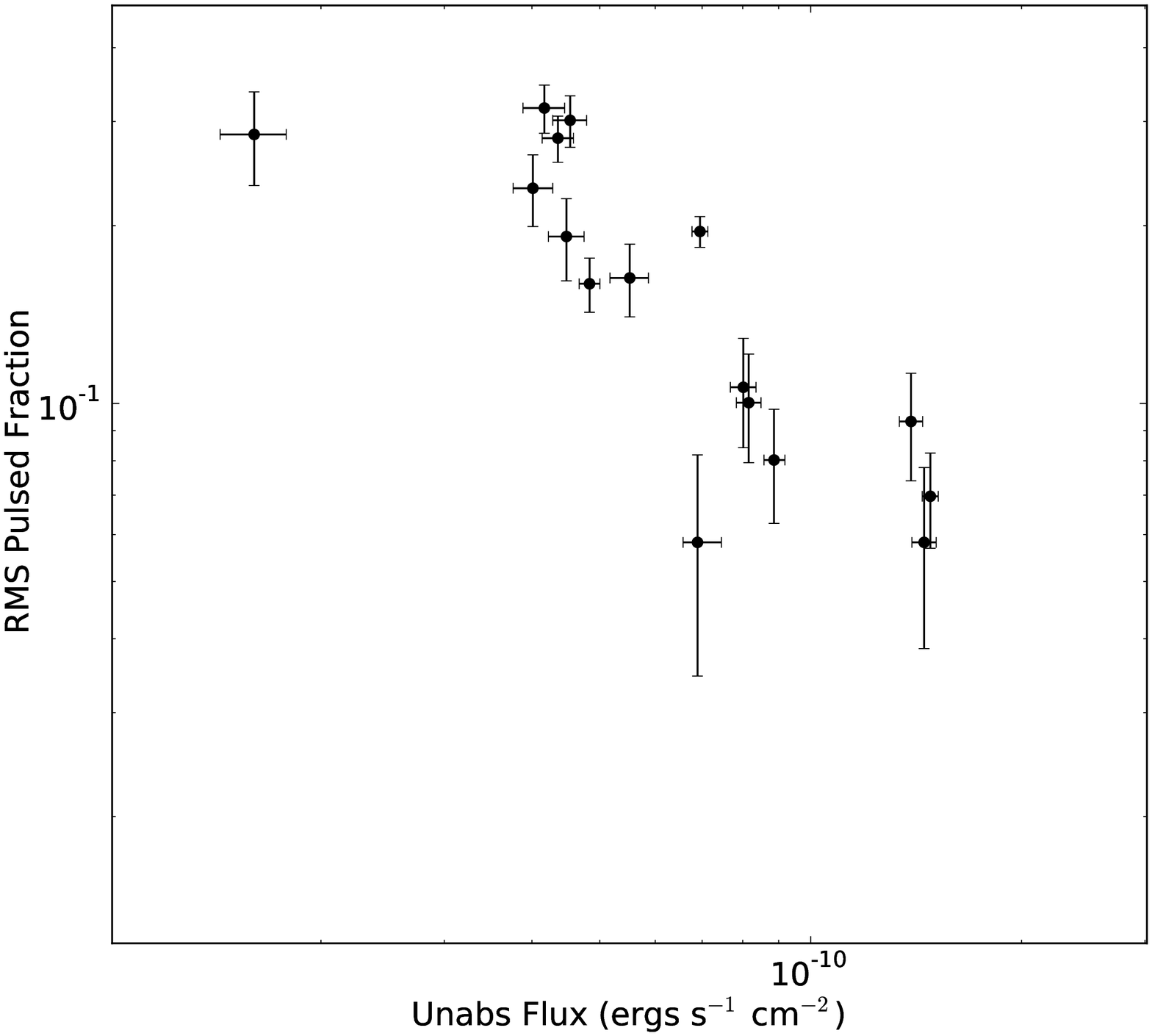}
\figcaption{1-10 keV RMS pulsed fraction as a function of 1-10 keV unabsorbed flux. 
 See also \citet{nkd+10}.
\label{fig:pfrac}
}
\end{figure}

Figure \ref{fig:pfrac} shows an anti-correlation between pulsed fraction and unabsorbed flux. It includes observations from both the 2008 and 2009
outbursts.
\citet{nkd+10} present this
trend using {\em Chandra} and {\em XMM} data as well as the burst-removed {\em Swift} data presented in this paper. 

\subsection{X-Ray Bursts} \label{sec:bursts}

Bursts were identified in a manner similar to that described in \citet{gkw04}. The event lists were binned at
a 1/16 s time resolution and a mean number of counts per bin was calculated for each Good Timing Interval (GTI). 
The number of counts in each bin, $n_i$, was compared to the GTI mean counts, $\lambda$, according to the probability $P_i$ of $n_i$ occuring randomly,
\begin{equation}
P_i = \frac{\lambda ^{n_i}e^{-\lambda}}{n_i!}.
\label{eq:poisson}
\end{equation}
Time bins that had $P_i \le 0.01/N$, where $N$ is the total number of time bins in the GTI being searched, were identified as part of a burst. 
Since the mean number of counts in a GTI can be overestimated due to contamination from bursts in other time bins, 
the procedure above was repeated iteratively, each time removing the bins that were identified as containing bursts, until no further bursts were identified. 
The above procedure was then repeated for 1/32 s and 1/64 s time resolutions to improve sensitivity to bursts of different durations. 

Since several bins identified with bursts can be part of the same burst, a burst was defined by its peak. The peak of a burst was 
determined by first finding the minimum time to accumulate 10 counts, using unbinned event data. The midpoint of the time spanned 
by these 10 counts was defined as the peak.
A search for a peak was done within 0.5 s on each side of an identified bin. This definition of burst peak is independent of binning and, 
for bursts with durations shorter than 1 s, will merge all of the identified bins into a single burst. Bursts within 1 s of each other that were
merged into a single burst were identified when selecting the background (see below) and their properties were measured separately. 

Once identified, to remove the bursts from the event lists, the lists were divided into full periods of the pulsar, starting with the first event of the  
observation as a reference point. The periods that contained bursts were identified and all counts which arrived in that interval were removed 
from the event list. This was done to ensure equal exposure to all pulse phases in the pulse profile. 

\subsubsection{Burst Statistics}

For each burst, a fluence, $T_{90}$, rise time ($t_r$), and fall time ($t_f$) were measured with the same analysis as in \citet{gkw04},
using the unbinned event data.
The fluence was determined by first measuring a background count rate in hand-picked regions on either side of the burst.
The background region by default was between 1 s and 2 s from the burst peak to either side of the burst, but was manually adjusted for most
bursts in order to avoid contamination from other nearby bursts.
The cumulative background-subtracted counts were then fit with a step function, using data point from the hand-picked background region.
The height of the step function in counts corresponds to the fluence of the burst.  The $T_{90}$ duration of the burst is the time between
when 5\% and 95\% of the fluence has been accumulated. The burst rise and fall times were determined using a maximum likelihood fit to a piecewise function
with an exponential rise and an exponential decay. Peak fluxes in counts per second were determined by passing a 62.5 ms boxcar integrator 
through a 250 ms interval (4 boxcar widths) 
centered on the burst peak. The highest count rate measured by the integrator was defined to be the peak flux. 

\begin{figure}
\plotone{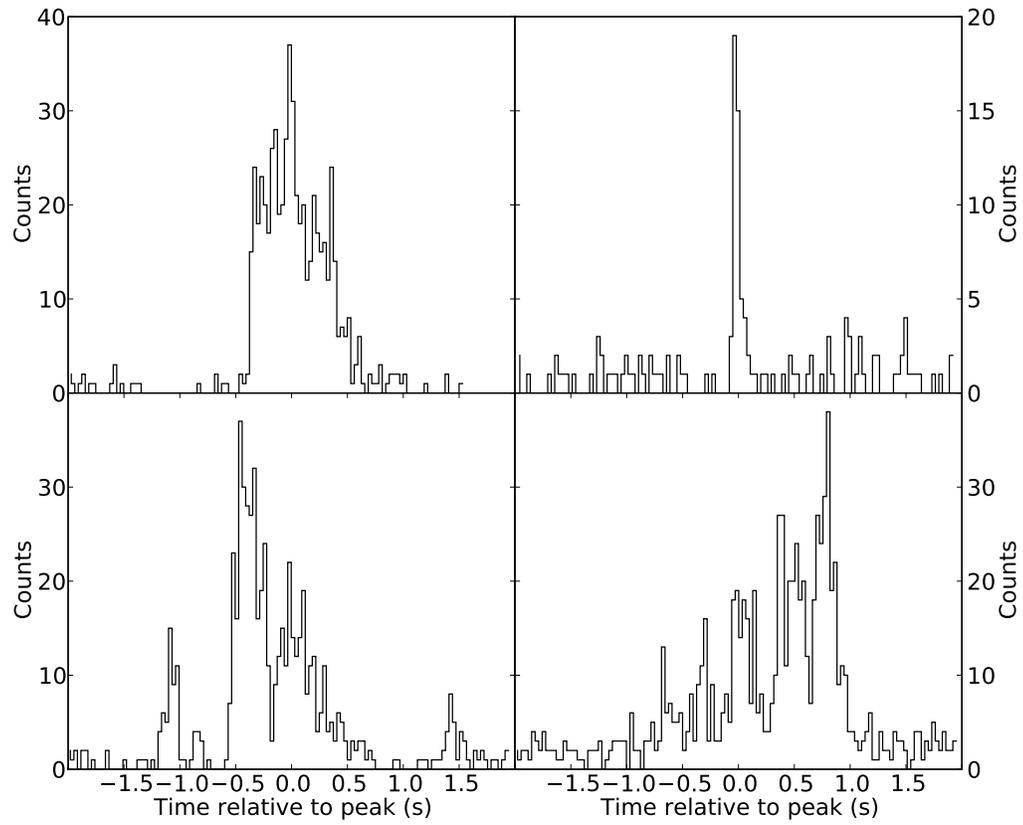}
\figcaption{Examples of bursts from 1E 1547$-$5408. Time series are binned on a 1/16 s timescale in the 0.5--10 keV energy range.
They display a wide range of morphology.
\label{fig:burstex}
}
\end{figure}

In total, for the 2009 outburst, 424 bursts were identified in 86 ks of observations from 2009 January 22 to 2009 September 30 using 1/16 s time resolution.
Thirteen additional bursts were identfied using the 1/32 s and 1/64 s time resolutions for a total of 437 bursts. 
Of those identified, 34 had too few counts to permit a reliable measure of fluence, 
and for 32, the exponential rise and decay were not successfully fit. For 64 of the bursts, properties could not be measured because they 
were too close to another burst to allow a reliable background estimate between them. Four bursts were too close to the edge of a GTI to calculate a 
background count rate. In summary, 303 bursts could be fully analysed and only these 2009 bursts will be considered henceforth. Examples of
different bursts are shown in Figure \ref{fig:burstex}.
Only two bursts were found in the XRT observations of the
2008 event; their properties were similar to those during the 2009 outburst. 

\begin{figure}
\plotone{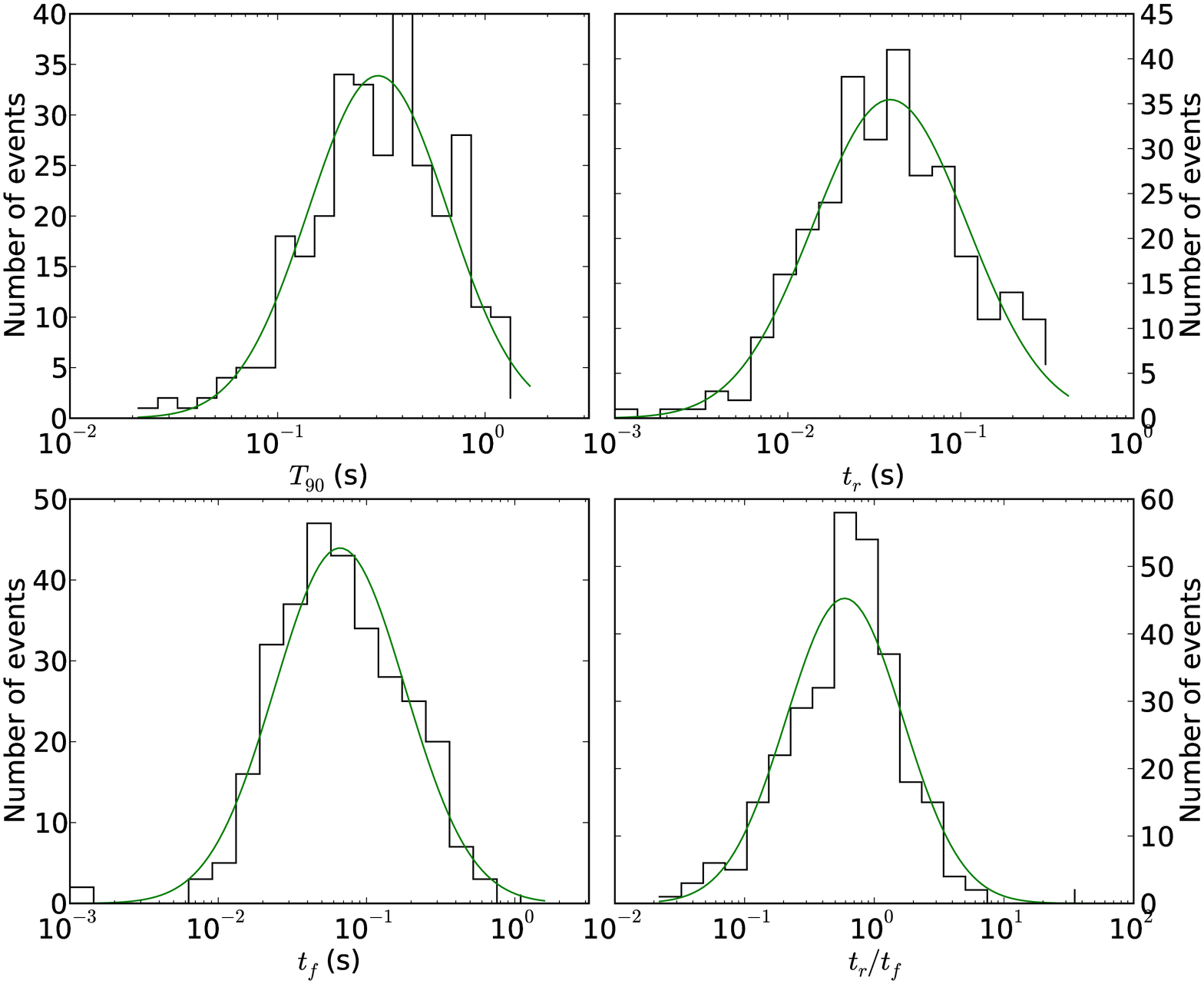}
\figcaption{{\em Top left:} Distribution of $T_{90}$ duration of bursts. {\em Top right:} Distribution of burst rise times.
{\em Bottom left:} Distribution of burst fall times. {\em Bottom right:} Distribution of $t_r/t_f$. In all panels, the 
solid line is the best-fit log-normal function.
\label{fig:burstprop}
}
\end{figure}

Figure \ref{fig:burstprop} shows the distributions of the burst properties, grouped in logarithmic bins. The $T_{90}$, rise time, fall time, 
and the ratio of rise to fall time distributions have been fit with log-normal distributions using maximum likelihood fitting. 
The $T_{90}$ distribution has a mean of 305 ms and a range for one standard deviation
of 140 - 662 ms. For the rise time distribution, we find a mean of 39 ms and a range for one standard deviation of 14 - 109 ms. 
For the distribution of fall times, we find a mean of 66 ms and a range of 24 - 182 ms for one standard deviation. The mean of the $t_r/t_f$
distribution is 0.59 with a range for one standard deviation of 0.21 - 1.66. A summary of the measured quantities is provided in 
Table 2. There we also provide these quantities, when available, for the three other sources for which such statistical analyses have been
done. We note in particular that the average burst duration for 1E~1547$-$5408 is longer than for all others measured thus far.
This result is further discussed in \S4.2.

\begin{figure}
\plotone{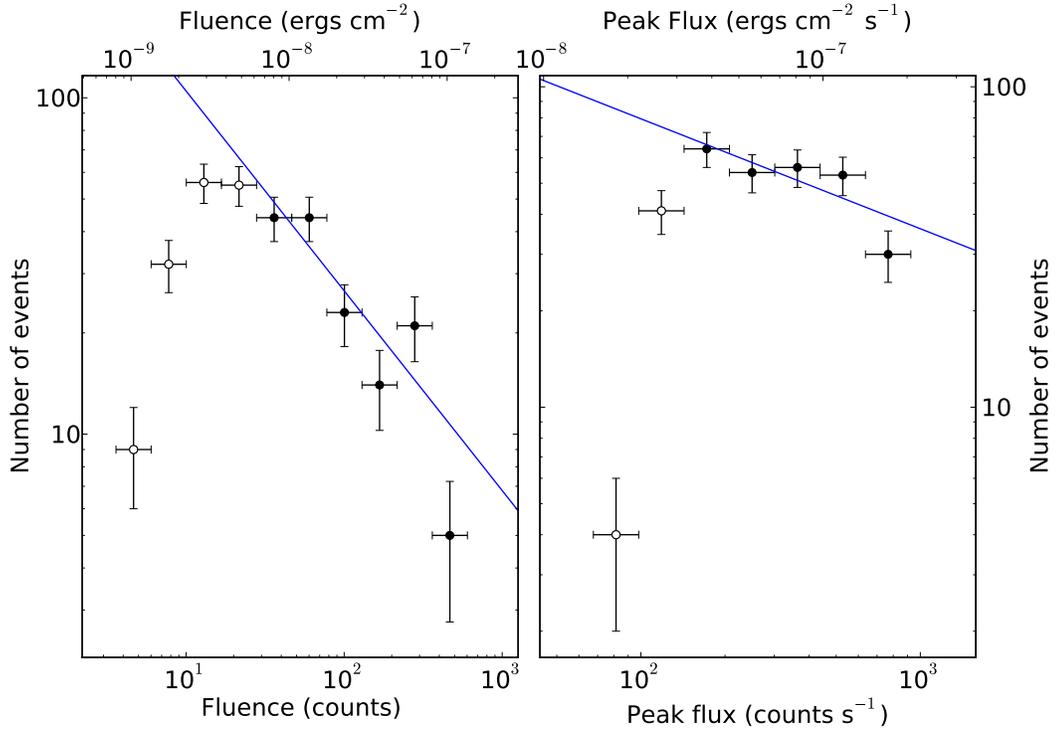}
\figcaption{{\em Left:} Distribution of burst fluences. {\em Right:} Distribution of burst peak fluxes. The solid line
is a linear fit to the filled circles. These distributions are based on burst counts from a 1--10 keV energy band.
The open circles are not included in the fits because of reduced sensitivity in
detecting such bursts. The top axes show the fluence and peak flux in cgs units which are derived from a single conversion
factor between counts and ergs cm$^{-2}$. In using such a factor, the bursts are assumed to have the same spectrum which is 
an approximation as each burst has a slightly different spectrum.
\label{fig:burstprop2}
}
\end{figure}

Figure \ref{fig:burstprop2} shows the fluence and peak flux distributions, in counts and counts per second, respectively, which have 
been fitted with a power law using a least-squares method. 
For low fluences or peak fluxes, the number is underestimated since the burst detection algorithm is less sensitive to these bursts. 
Therefore, some of the points with low fluence or peak flux were not included in the fits. The best-fit power-law index for the fluence distribution is 
$-0.6 \pm 0.1$. For the peak flux distribution, the power-law index is $-0.34 \pm 0.12$. This index is quite shallow and, despite the omission of the 
first two points in the fit, may still be affected by the bias in the burst search.

\begin{figure}
\plotone{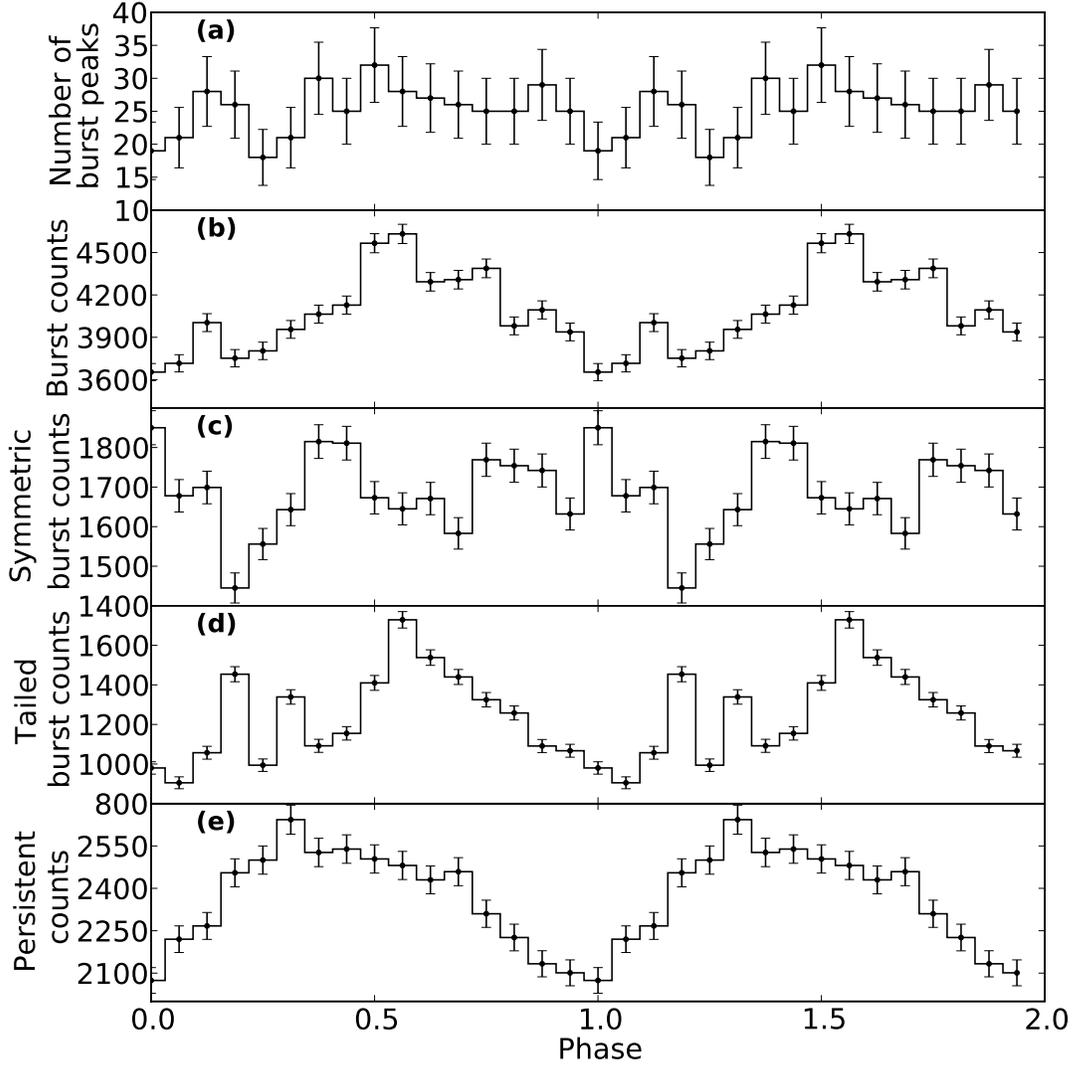}
\figcaption{(a) Folded profile of the times of the burst peaks. (b) Folded profile of photon
counts from cycles of the pulsar that contain bursts. (c) Folded profile of photon counts from cycles containing symmetric bursts.
(d) Folded profile of photon counts from cycles containing slow-fall bursts. (e) Folded profile of photon
counts from cycles of the pulsar that do not contain bursts but not including the
first two observations following the BAT trigger, illustrating the quiescent pulse profile. Profiles are all in the 0.5--10 keV energy range.
\label{fig:burstprof}
}
\end{figure}

For other AXPs, there is evidence that bursts tend to arrive on pulse \citep[e.g.][]{gkw02, gkw04}. For 1E 1547$-$5408 this does not seem to be the case, as burst peaks are 
distributed randomly in phase (see Fig. \ref{fig:burstprof}a). This is similar to what is observed in SGRs 1806$-$20 and 1900+14 \citep{pal99,pal02}. 
However, when the individual photon arrival times that are part of bursts are folded, a strong pulse is 
observed. The peak of this `pulse', shown in Figure \ref{fig:burstprof}b, is not aligned with the peak of the quiescent pulse profile. 
Figure \ref{fig:burstprof}e presents this quiescent pulse profile obtained using the burst-removed data. 
Since the pulsed fraction of the first two observations following the 2009 BAT trigger is significantly lower and the persistent flux is significantly higher than in the
subsequent observations, including them in the profile reduces the pulse amplitude. Thus, in order to better show the persistent pulse profile, 
Figure \ref{fig:burstprof}e does not include those first two observations.
In order to determine whether the burst count pulse was dominated by a handful of bright bursts, the 15 brightest bursts were removed and the counts were refolded. 
The profile did not change significantly and still displayed a pulse at a similar phase.

Panels c and d of Figure \ref{fig:burstprof} separate the symmetric bursts from slow-fall bursts. Symmetric bursts were defined as bursts with $0.5 < t_r/t_f < 2$.
Bursts with $t_f$ greater than $2\:t_r$ were defined as slow-fall bursts. 
Of the 303 well measured bursts, 158 were classified as symmetric and 116 as slow falls. The remainder, 29 bursts,
were those with rise times greater than twice their fall times. 
Curiously, the folded slow-fall burst counts exhibit a much stronger pulse than do the folded symmetric burst counts,
with a $\chi^2_\nu = 43$ for the null hypothesis, compared to $\chi^2_\nu = 6$ for the folded symmetric bursts.
This symmetric/slow-fall definition is somewhat arbitrary and these two classes of bursts are by no means distinct 
populations. We use this distinction only to demonstrate that the more symmetric bursts tend to be more pulsed than those that are less symmetric.

\subsubsection{Burst Spectroscopy}

Spectra within the $T_{90}$ interval of each burst peak were extracted and grouped with 20 counts per bin. Background spectra were taken from 1 min on either side
of the burst peak extracted from the burst-removed event data. The 46 bursts with fluences over 200 background-subtracted counts were fitted with a 
photoelectrically absorbed power law with $N_H$ fixed to $3.24 \times 10^{22}$ cm$^{-2}$ as measured from the fit to the persistent emission (see \S3.1). 
The spectra were fitted using XSPEC. 
The mean of the measured spectral indices of the bursts was found to be 
$\Gamma = 0.17$, with a standard deviation of 0.33, where $ N(E) \propto E^{-\Gamma}$.
Fitting with more complicated models was attempted, but parameters could not be successfully constrained because of the low number of counts. 
As shown in Table 2, the average $\Gamma$ we measure is significantly harder than that measured in the only other magnetar outburst 
for which the value is reported.  This is discussed further in \S4.2.

In order to determine the effect of pileup on the bursts, power-law spectra were fit to the two bursts with the highest peak flux while excluding a central region
ranging from 1 to 15 pixels in radius. The power-law indices from the fits were plotted against radius of exclusion region as per the prescription of \citet{rcc+06}.
The power-law index did not change significantly as a function of exclusion radius. Hence we concluded that the bursts were 
not significantly affected by pileup.  

\begin{figure}
\plotone{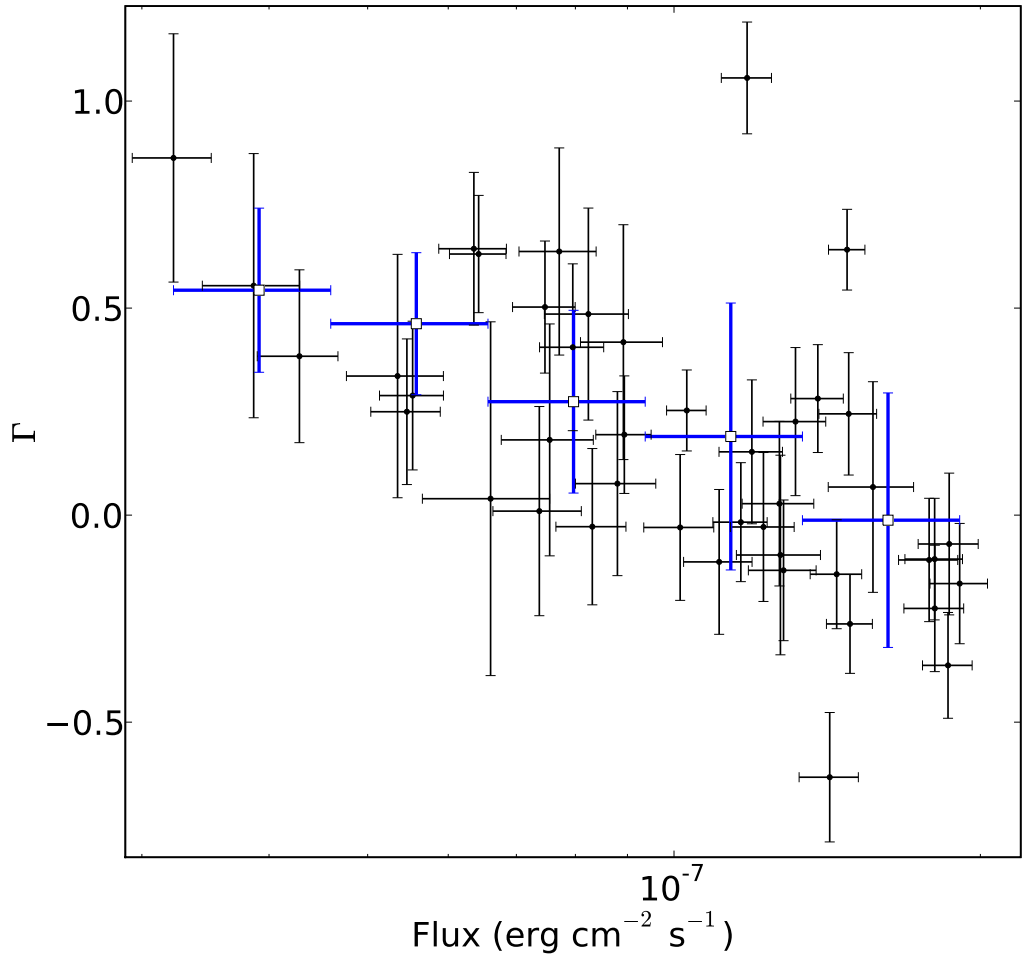}
\figcaption{ Power-law index as a function of average absorbed flux over $T_{90}$ from the spectral fits of the 46 most fluent bursts. The black points are the individual bursts 
and the blue open squares represent the weighted averages of the power-law indices for bursts in logarithmic fluence bins.
\label{fig:fluxgamma}
}
\end{figure}

In order to probe the relation between hardness and fluence for the bursts, hardness ratios were determined for each burst by measuring fluences 
in the 0.5--4~keV and 4--10~keV bands. 
No significant correlation between 0.5--4~keV/4--10~keV hardness and fluence was observed for bursts from 1E 1547$-$5408. For the 46 most fluent bursts, 
there was also no observed correlation between the 0.5--4~keV/4--10~keV hardness or the spectral index of the power-law fit and the fluence. 
The lack of a detection of the correlation between hardness and fluence in the 
{\em Swift} data may be due to the limited spectral range of the XRT (0.5--10 keV).
However, an anti-correlation between power-law index and average flux over $T_{90}$ from the spectral fits was observed 
(Fig. \ref{fig:fluxgamma}). This is consistent with a correlation between the hardness and the magnitude of a burst.

Spectral features in magnetar bursts have been reported for some sources \citep{si00,iss+02,gkw02,isp03,wkg+05,gdk09,ks10}, although most of the features have been
discovered above 10 keV. The individual burst spectra for 1E 1547$-$5408 
showed no evidence for any spectral features in the 1-10 keV range. Since spectral features may be masked by the low count rates in the individual spectra, 
we combined the burst spectra into a single average spectrum. The burst and background spectra and response filestend to bere combined using the FTOOL {\ttfamily addspec}. 
Spectral bins were grouped with a minimum of 20 counts per bin. The average spectrum was fit with a blackbody, a power law, a blackbody with an added power law, 
and a Comptonized blackbody; all four models were photoelectrically absorbed. All of the models gave acceptable fits with $\chi^2_\nu \sim 1$;
thus, no significant spectral features were observed in any of the residuals.

Fluences and peak fluxes in counts and counts per second, respectively, were converted into cgs units using the results of the spectral fits. 
The fluxes from the power-law fits were multiplied by the burst durations
and compared to the fluences measured in counts. 
The proportionality constant between the two was determined to be $2.23 \times 10^{-10}$ ergs cm$^{-2}$ counts$^{-1}$. 
In using this single conversions factor between counts and ergs cm$^{-2}$ we are effectively assuming an average spectral model
for the bursts which is not generally true. The conversion from counts to cgs units in reality is different from burst to burst
depending on their spectra.  
Values in cgs units, using this single conversion factor, are indicated on the top axes of the 
fluence and peak flux distributions in Figure \ref{fig:burstprop2}.

\section{Discussion}

We have presented {\em Swift} XRT observations of the 2008 and 2009
outburst events of 1E 1547$-$5408.
In particular, we considered the behavior of this source's persistent flux in
its 2009 outburst, which was qualitatively different from that in 2008.
Specifically, the 2009 event showed a much smaller increase in pulsed flux
than in 2008, while the opposite is true of the total flux.  In this work,
we have shown that for the 2009 outburst, the source spectrum became harder 
(though the exact degree of hardening is presently difficult to know, due to the contaminating dust scattering)
as the flux increased after the initial trigger, and we found an anti-correlation between the pulsed
fraction and the flux.

We have also presented a detailed study of the several hundred X-ray bursts
detected by {\it Swift} following the 2009 event.  Noteworthy results we have
found include
a correlation between burst flux and hardness.  We also note
that burst counts tend to contribute very significantly to the pulsed
fraction, even if the phases of burst peaks are randomly distributed
in pulse phase.

Next we discuss our results in the context of the magnetar model, as well as in
comparison with those for other similar outbursts.

\subsection{Persistent Flux}

Figure \ref{fig:spec} shows that as the X-ray flux increased, the blackbody
temperature increased and the power-law index decreased during the
2009 outburst.  Such a hardness/flux correlation was seen also in the 2008 outburst
\citep{ier+10} and ubiquitously in other magnetar outbursts
\citep[e.g.][]{kgw+03,wkt+04,icd+07}.
In the case of 1E 1547$-$5408, however, we regard the precise
values of $kT$ and $\Gamma$ as shown in Figure \ref{fig:spec} on the first day after the outburst with caution, as they
are very likely biased by dust scattering.  
The hardness/flux correlation for this outburst on the day of the trigger is likely to be robust, since dust scattering softens photons. 
However, note that if the dust scattering rings are caused by a burst that is much harder than, and separate from, 
the persistent emission, the dust scattering could effectively cause the overall emission to 
harden in relation to the true persistent emission.

The thermal emission of magnetars, in the twisted magnetosphere
model \citep{tlk02}, has as its origin heating from within the star, due
to the decay of the strong internal magnetic field.  The resulting
thermal surface photons are thought to be
scattered by currents in the atmosphere, resulting in a
Comptonized blackbody-like spectrum which is often modelled with a
blackbody plus a power law \citep{lg06}.  The magnetospheric currents are present
due to `twists' in the field structure, either global \citep{tlk02},
or, more likely, in localized regions \citep{bel09}.
In either case, the current strength, hence degree of scattering, 
increases with increasing twist magnitude.
Moreover, return currents provide an additional, external source of
surface heating in addition to the internal source.
The increase in flux during a magnetar outburst is thus
theorized to be caused by an internal heat-releasing event
that may significantly increase the surface temperature \citep[see, for
example, ][]{og07}, further twist the magnetospheric field, and increase
the external return-current heating.  Thus, a correlation
between hardness and flux is generically expected in the twisted-magnetosphere model \citep{lg06},
in agreement with observations of 1E 1547$-$5408 and other magnetars.

\begin{figure}
\plotone{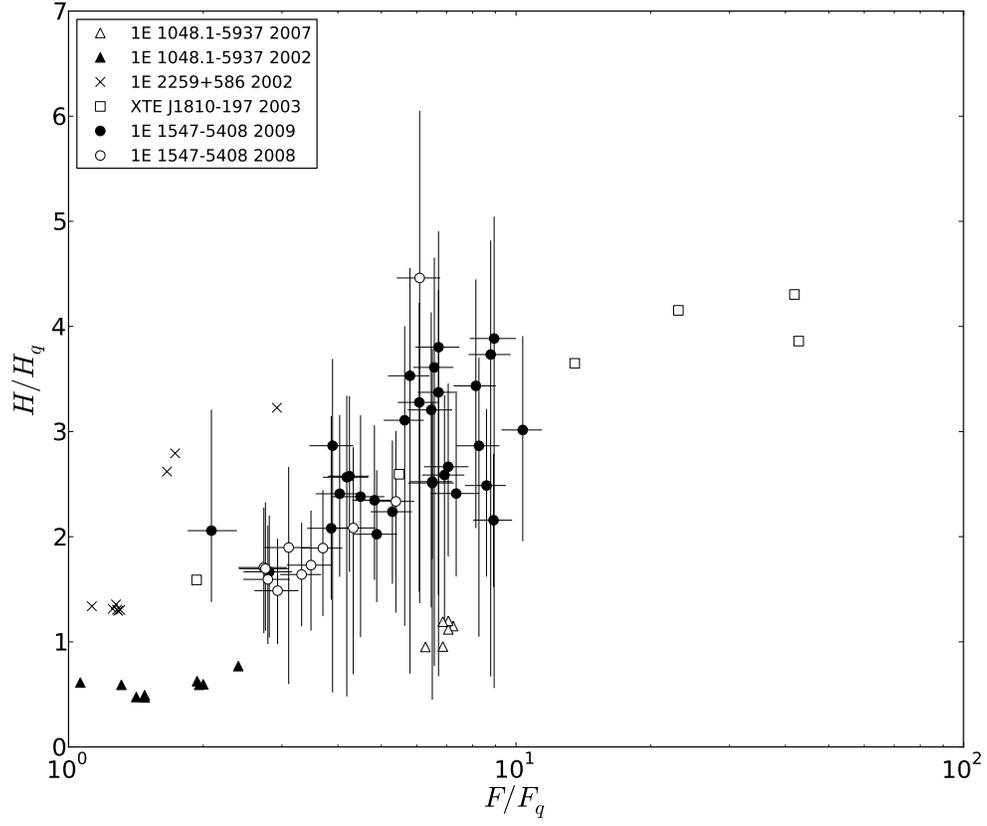}
\figcaption{4--10 keV / 2--4 keV flux hardness, $H$, as a function of 2--10 keV flux, $F$ for magnetar outbursts.
Both are normalised to their quiescent values, $H_q$, $F_q$.
For 1E 1547$-$5408, the spectral fits used to determine the hardnesses and fluxes were from this
work, and the first day of the 2009 outburst has been omitted due to contamination by the dust-scattered emission. 
For XTE J1810$-$197 the spectral parameters were taken from \citet{gh07a}.
For 1E 1048.1$-$5937 they were taken from \citet{tgd+08}. For 1E 2259+586 they were taken from \citet{zkd+08}. 
Uncertainties on the hardness ratios are shown only for 1E~1547$-$5408; determining those for other
sources is difficult from the literature, however they are likely comparable to the scatter. 
\label{fig:hardening}
}
\end{figure}

It is interesting to compare the hardness/flux correlations
seen for 1E~1547$-$5408 with those observed for other magnetars to see if there exists
a universal relationship between increase in X-ray flux and hardness,
as might be expected in the above scenario.  Figure~\ref{fig:hardening} 
shows the fractional increase in the 4--10/2--4~keV
hardness ratio determined from fluxes as a function of 2--10~keV unabsorbed flux fractional
increase for six different magnetar outbursts in four different magnetars. 
We chose to consider a flux hardness ratio as a measure of hardness
(rather than e.g. power-law index) as it is defined independent of spectral model,
and also is instrument-independent.
In order to determine the hardness of each source, spectral model
parameters from the literature were compiled and input to XSPEC. The 4--10~keV and 
2--4~keV fluxes were then determined using the XSPEC {\tt flux} command and
the hardness was calculated from their ratio. For references that did
not report a 2--10 keV unabsorbed flux, one was determined using XSPEC using the reported model paramenters.
As is clear from the Figure, an overall trend is apparent, although there
is significant variation from source to source.  For example, for similar
flux enhancements over the quiescent level, 1E~2259+586 became much harder
than did 1E~1048.1$-$5937, with the behavior of 1E~1547$-$5408 lying somewhere
in between.  Thus, though
Figure~\ref{fig:hardening} demonstrates that a hardness/flux
correlation is indeed generically observed, there does not appear to exist a universal
law linking the degree of flux increase over the quiescent level with the degree of flux
hardening.  This observation will require accommodation
in any detailed model of magnetar outburst emission.

As shown in Figure \ref{fig:pfrac}, the flux of 1E 1547$-$5408
was anti-correlated with the pulsed fraction.  While immediately
preceding the 2009 outburst the pulsed fraction was over 20\%
(Fig. \ref{fig:spec}), immediately afterward, it dropped to nearly
negligible levels in the 1--10 keV range, likely due to the emission
being dominated by the scattered emission from the dust scattering rings.
However, following the first day, the pulsed fraction was still significantly
lower than the preburst pulsed fraction.
This anti-correlation between pulsed fraction and flux is presented
with the addition of {\em Chandra} and {\em XMM-Newton} data
in \citet{nkd+10}, who suggest it could be due to an
increase in the emitting area of a thermal hot spot, which would cause the
emission to be observable during a greater portion of the phase of the
pulsar. We note that both correlations \citep{ghbb04,zkd+08} and anti-correlations
\citep{icd+07,tgd+08} between pulsed fraction and flux have been
observed in magnetar outbursts.  Different behaviors
can result depending on the location of the
emission region and the viewing geometry.  Although detailed modeling of spectral,
pulse morphology and pulsed flux changes during magnetar outbursts is
beyond the scope of this paper, recent attempts at unified modelling of
magnetar surface and magnetosphere emission geometries and emission
mechanisms \citep{ati+10} could in principle use data like ours to constrain the hot spot
location and size.  This seems to us to be a good future avenue for investigation.

Interestingly, \citet{kgk+10} report a $\sim$55\% pulsed fraction
at $\sim$100 keV as observed by {\em Fermi} GBM $\sim$30--40 minutes before
the BAT trigger. They
report that this is the first time that pulsations unrelated to a giant
flare have been observed at $\sim$100 keV for an SGR\footnote{Note
that \citet{kgk+10} use the designation SGR 1550$-$5418 for 1E
1547$-$5408.}. However, AXPs have previously been shown to exhibit
pulsations at $\sim$100 keV with pulsed fractions as high as 100\%
\citep[e.g.][]{khhc06}.  Thus it is possible that the measured
55\% actually decreased from a higher pulsed fraction leading up to the
outburst event, i.e. the $\sim$100~keV pulsed fraction may have behaved
similarly to that in the 1--10 keV band.  Future hard X-ray telescopes
with focusing optics, such as {\em NuSTAR}, will allow much easier
measurements of the pulsed fraction of magnetars at high X-ray energies.


\subsection{Bursts}

Previous studies of SGR bursts have shown that their energies, and thus fluences, follow a
power-law distribution $dN/dE \propto E^{-\alpha}$ with $\alpha$ equal
to $\sim$5/3 \citep{cegy96,gwk+99,gwk+00}. It has been noted that this
is similar to the Gutenberg-Richter law for earthquakes and to energy
distributions of solar flares \citep{cad93,lhmb93}.  Both 1E 1547$-$5408
and 1E 2259+586 also follow this distribution. In this work, the fluence
distribution of 1E 1547$-$5408 is found to have a power-law index of
$-0.6\pm 0.1$ which corresponds to $dN/dE \propto E^{-1.6}$. 
\citet{gkw04} find an $\alpha$ of $1.7\pm0.1$ for 1E
2259+586, similar to the values found for SGRs, further reinforcing
the similarity in this particular behavior among AXPs and SGRs. 

For the 2009 outburst, {\em INTEGRAL} observations which cover an
energy range of $>80$ keV, show different burst properties from those determined
using the {\em Swift} XRT observations. \citet{snb+10} report a 68-ms mean
duration derived from a log-normal distribution with a scatter of 30 - 155 ms. 
This is much shorter
than the 305-ms duration determined in this work (see Table \ref{ta:mags}).  This discrepency may
be due to the difference in energy coverage; perhaps
bursts have different morphologies at different energies.  However, the
definition of duration used by \citet{snb+10} differs from the $T_{90}$
definition used here.  \citet{snb+10} define the burst duration as
the time between the moment when the count rate rises above 5$\sigma$
to when the count rate drops below 3$\sigma$.
When applying their definition of duration to the bursts identified
in our study, we find, for a log-normal distribution, a mean duration of 101 ms 
and a range for one standard deviation of 59 - 173 ms, closer to but 
still somewhat longer than their measurement, suggesting a possible energy dependence
of burst duration.  We do not, however, detect any significant difference
in the durations measured using 0.5--10 keV counts with those measured
using 2--10~keV counts.


The properties of bursts from the 2009 outburst event of 1E 1547$-$5408
are reminiscent of those from the outburst of 1E 2259+586 \citep{gkw04}. Both
have a significant number of short spikes like those found in SGRs,
and a set of bursts with long pulsating tails like those found in burst
studies of other AXPs. Although we do not find any bursts with long pulsating tails in the XRT observations,
\citet{snb+10} and \citet{mgw+09} find two such bursts. These were not found in our analysis because {\em
Swift} was not observing 1E 1547$-$5408 when they occured. 
Table \ref{ta:mags} compares the properties of bursts from 1E~1547$-$5408 to those from outbursts from other magnetars.
We note that the average durations of bursts from 1E~1547$-$5408 appear to be
longer than for those in other sources. However, this could be an artifact as
the burst properties for the other tabulated sources were determined with {\em RXTE}.
The larger collecting area of {\em RXTE} allows it to detect bursts that are fainter 
than those {\em Swift} can detect. If faint, short bursts are missed by the XRT,
the mean burst duration (as well as $t_r$ and $t_f$) may be overestimated.
Also, the energy range of RXTE (2--60 keV) probes higher energies and so 
differences may be due to the energy dependance of burst properties.
A detailed statistical study of 1E~1547$-$5408 bursts with {\em RXTE} is
needed to clarify this point. 

Although {\em Swift} XRT is not ideal for probing the spectra of magnetar bursts, which have significant flux above 10 keV,
we were still able to draw the following conclusions from our analysis.
While we do not observe a hardness-fluence
correlation for 1E 1547$-$5408, this could be due to our limited
energy range. Indeed, there is a hint of a correlation between $\Gamma$ and fluence, however
is is not statistically significant. Note however that we do observe a significant $\Gamma$-flux correlation
(Fig.~\ref{fig:fluxgamma}). 
\citet{snb+10}, using {\em INTEGRAL} data of the 2009 outburst, find a correlation between burst
hardness and count rate. Their hardness ratio is defined as the ratio between the Anti-Coincidence Shield (ACS) flux, which is sensitive to photons above 80 keV, 
and a 20-60 keV flux from the ISGRI instrument. This is also consistent with a correlation
between hardness and burst magnitude for 1E 1547$-$5408. 
\citet{gkw04} also find a correlation between hardness and fluence for 1E 2259+586.  
For the SGRs, on the other hand, an anti-correlation between
hardness and fluence has been observed \citep{gwk+99,gwk+00}. 
Table \ref{ta:mags} also shows that the 1E~1547$-$5408 
bursts from {\em Swift} are much harder than those from 1E~2259+586. In light of the observed hardness-fluence 
correlation in AXP bursts, this is not a suprising result. The 28 most fluent bursts from 1E~2259+586 for
which spectral indices were measured, have fluences of  $\sim10^{-9} - 10^{-8}$ erg cm$^{-2}$ \citep{gkw04}.
This is to be compared with the 46 most fluent bursts for which $\Gamma$ was measured here for 1E~1547$-$5408, that span a fluence range 
of $\sim10^{-8} - 10^{-7}$ erg cm$^{-2}$. This may account for the harder average spectral index for bursts from 1E~1547$-$5408.



In the magnetar model, two mechanisms have been suggested for producing magnetar
bursts. \citet{td95} suggest that stresses due to the strong magnetic fields present
inside magnetars are able to crack the crust of the neutron star. This
cracking releases a plasma fireball into the magnetosphere. The strong
magnetic fields can hold the fireball above the fracture site. The suspended
fireball can heat the surface thus causing an extended cooling tail. Since
the strongest surface fields are located near the polar caps, these
fracture events would occur preferentially near the poles which would
result in an observed phase dependence of the bursts.
Another mechanism, proposed by \citet{lyu03}, suggests that bursts
are caused by reconnection events initiated by the development of a tearing mode
instability in the magnetically dominated relativistic plasma in the magnetosphere. In this case,
bursts occur randomly in
phase. This mechanism should also produce shorter more symmetric
bursts than in crustal fracture. \citet{lyu03} states that
the hardness-fluence anticorrelation found in SGRs is consistent with
magnetic reconnection. For the surface-cooling model, the opposite is
expected.  As discussed by \citet{lyu03}, both mechanisms could easily
be at work.

\citet{wkg+05} suggested that observationally, there are two types of magnetar bursts.
Type A bursts are nearly symmetric and are typical
of SGR bursts.  Type B bursts have been observed in AXPs and are
characterized by a short spike followed by a long tail, typically much
longer than the rotational period of the pulsar. Pulsations have been
observed in these tails. Since the Type B bursts observed in AXPs occur
preferentially in phase and exhibit a hardness-fluence correlation, and
the Type A bursts in SGRs are distributed randomly in phase and have a
hardness-fluence anti-correlation, \citet{wkg+05} suggest the magnetic
reconnection mechanism for type A bursts and the surface-cooling model
for Type B bursts.

For 1E 1547$-$5408, none of the bursts detected by {\em Swift} XRT could be
classified as Type B. However, the two bursts with long pulsating tails found in {\em INTEGRAL} 
data by \citet{snb+10} and \citet{mgw+09} have typical Type B morphology.
For the bursts in this work, over half of the bursts were classified as
symmetric, which nominally corresponds to Type A (see top-right panel of Figure
\ref{fig:burstex} for example). About 40\% of the bursts were classified
as slow-fall bursts, which are actually closer to Type A in morphology than Type B,
although they are not very symmetric.   
Thus, the bursts from 1E~1547$-$5408 are not easily classified into
Types A and B as described by \citet{wkg+05}. Moreover, for both the symmetric and
slow-fall bursts, the folded photon arrival times exhibited a clear phase dependence, 
although for the slow-fall bursts this `pulse' is much stronger. That the
symmetric bursts show some pulse modulation is suggestive of a difference with `classical'
Type A bursts, or that pulse phase analyses of the latter should be attempted using all burst
counts, as they too may be pulsed. 
That the burst counts are pulsed indicates that burst emission comes from
a preferred region in rotational phase, be it on or near the surface or high in the magnetosphere,
even if the burst peaks arrive randomly in phase.  The offset between the burst counts pulse
peak and the persistent pulse peak seen when comparing Figures 7b and 7e demonstrates that
the preferred burst emission region has location and geometry similar to, but distinct from,
that producing the persistent pulsations.



The total energy released in the {\em Swift}-detected bursts was $\sim
1 \times 10^{40}$ erg in the 1-10 keV range.  This is much lower than
the 1-10 keV energy released from the persistent emission between 2009
January 22 and 2009 September 30 of $\sim 9 \times 10^{41}$ erg. For
reference, the energy released from the spin-down in that same period
is $\sim 5 \times 10^{40}$ erg. \citet{wkt+04} find that for SGRs, the
energy released in the bursts is higher than that released in the
persistent emission, but for 1E 2259+586, the opposite is true.
In this regard, the 2009 outburst of 1E~1547$-$5408 is more like that
of AXP 1E~2259+586.


\section{Conclusions}

We have presented an analysis of the persistent radiative evolution of the 2009 January outburst of 1E~1547$-$5408 
from {\em Swift} XRT observations. We found that $\sim6$~hr after the 
2009 BAT trigger the observed persistent 1--10~keV unabsorbed flux reached a peak of  $\sim8 \times 10^{-9}$ ergs cm$^{-2}$ s$^{-1}$, an increase of more 
than 500 times the quiescent flux leading up to the outburst. This flux evolution is not due solely to the source;
in the first day, there is also emission from dust scattering rings that is delayed emission from 
an energtic event near the onset of the outburst.
There was significant spectral hardening at the outburst as seen in 
other magnetar outbursts. 
Note that the absence of spectral variation reported by \citet{nkd+10} is consistent with our results, as they missed the bulk of the
spectal changes which occured in the first day of the outburst and their {\em Chandra} observations did not begin until the next day.
The pulsed fraction showed an anti-correlation with the 
phase-averaged flux for both the previous 2008 and 2009 outbursts, with both sets of data following the same trend. 
We have compiled data from this and five other magnetar outbursts for four different sources and find
a generic X-ray hardness/flux correlation overall, but with no clear universal quantitative relationship between the two.

We have also presented a detailed statistical analysis of the several hundred bursts detected during the 2009 event. 
The bursts do not easily fall into the Type~A/Type~B classification put forth by \citet{wkg+05}.
We found that the peaks of the bursts were
randomly distributed in pulse phase, but that when the individual photon counts were folded at the pulsar emphemeris, a clear
pulse was present. This phase dependence is stronger for those bursts with longer decays than rise times than for
those bursts that are symmetric. We also report a correlation between burst hardness and burst flux. 

In many ways, these observations yield more questions than answers.  The range of
observed phenomenology in magnetar outbursts seems to increase with each event, with
few overall trends to assist in constraining models emerging.   Nevertheless given the paucity
of events studied in detail, we remain hopeful that through perserverence, eventually physical
insight will emerge.   The fast response of telescopes like {\it Swift} is crucial to this
endeavor.  For 1E 1547$-$5408, it allowed an analysis of the first day of the 2009 event.
This is important as both the most significant spectral changes and the majority of the bursts occured within
this period. This highlights the necessity of prompt response to magnetar outbursts in understanding their nature.
\\

After submission of our manuscript, we became aware of the work by \citet{bis+11} on the 2009 outburst of 1E 1547$-$5408
using {\em Swift} XRT data.
A preliminary comparison of their results with ours shows general
agreement.  In particular their reported overall source behavior, namely spectral hardening
correlated with flux, as well as pulsed fraction anti-correlated with flux and
blackbody radius, are in broad agreement with our results.  
We further note that they omitted reporting
on data taken with {\em Swift} XRT in the 24 hrs following the 2009 trigger, due to the difficulty
in handling the dust-scattered emission (P. Esposito, A. Tiengo, private communication).
\\


We are grateful to P. Esposito and A. Tiengo for sharing with us their preliminary work on
modelling of the dust-scattered emission on the first day following the onset of the 2009 outburst.
We thank A. Archibald, P. Lazarus, C.-Y. Ng, and S. Olausen for useful discussions. We thank R. Dib for providing an
ephemeris and pulsed count rate measurements from {\em RXTE}. We thank P. Woods for helpful comments.
V.M.K. holds the Lorne Trottier Chair in Astrophysics and Cosmology and a Canadian
Research Chair in Observational Astrophysics. This work is supported by NSERC via a Discovery Grant, by FQRNT, by CIFAR,
and a Killam Research Fellowship.

\bibliographystyle{apj}
\bibliography{journals_apj,myrefs,modrefs,psrrefs,crossrefs}

\begin{table}[t]
\begin{center}
\caption{Summary of {\em Swift} XRT observations of the 2009 outburst of 1E 1547$-$5408}
\begin{tabular}{ccccccc} 

\hline\hline
Sequence    & Mode    & Observation date & MJD     & Exposure time       & Time since trigger \\
            &         &                  & (TDB)   &   (ks)               &  (days)          \\

\hline

00090007024 & WT & 2009-01-04 & 54835.0 & 3.3 & $-18.033$ \\
00090007025 & WT & 2009-01-12 & 54844.0 & 4.2 & $-9.066$ \\
\hline
00340573000 & WT & 2009-01-22 & 54853.1 & 6.1 & 0.035 \\
00340573001 & WT & 2009-01-22 & 54853.4 & 9.2 & 0.321 \\
00340923000 & PC & 2009-01-23 & 54854.6 & 1.7 & 1.575 \\
00090007026 & WT & 2009-01-23 & 54854.7 & 8.2 & 1.657 \\
00340986000 & PC & 2009-01-24 & 54855.2 & 2.9 & 2.110 \\
00030956031 & PC & 2009-01-24 & 54855.3 & 2.5 & 2.195 \\
00090007027 & WT & 2009-01-25 & 54856.1 & 3.3 & 2.990 \\
00341055000 & PC & 2009-01-25 & 54856.1 & 4.0 & 3.065 \\
00341114000 & PC & 2009-01-25 & 54856.9 & 4.6 & 3.851 \\
00090007028 & WT & 2009-01-26 & 54857.1 & 3.5 & 3.987 \\
00030956032 & PC & 2009-01-27 & 54858.1 & 6.2 & 4.996 \\
00090007029 & WT & 2009-01-27 & 54858.4 & 1.8 & 5.330 \\
00030956033 & PC & 2009-01-28 & 54859.2 & 5.1 & 6.143 \\
00090007030 & WT & 2009-01-28 & 54859.9 & 1.9 & 6.811 \\
00030956034 & PC & 2009-01-29 & 54860.0 & 5.9 & 6.941 \\
00090007031 & WT & 2009-01-29 & 54860.7 & 2.2 & 7.614 \\
00090007032 & WT & 2009-01-30 & 54861.5 & 2.9 & 8.476 \\
00030956035 & WT & 2009-01-30 & 54861.7 & 3.0 & 8.677 \\
00030956036 & WT & 2009-01-31 & 54862.1 & 3.0 & 9.076 \\
00090007033 & WT & 2009-01-31 & 54862.8 & 2.5 & 9.687 \\
00030956037 & WT & 2009-02-01 & 54863.5 & 2.0 & 10.476 \\
00090007035 & WT & 2009-02-02 & 54864.6 & 4.1 & 11.548 \\
00030956038 & PC & 2009-02-03 & 54865.5 & 5.9 & 12.424 \\
00030956039 & PC & 2009-02-04 & 54866.7 & 6.1 & 13.637 \\
00030956040 & WT & 2009-02-05 & 54867.2 & 6.1 & 14.101 \\
00030956042 & WT & 2009-02-07 & 54869.3 & 1.7 & 16.189 \\
00090007036 & WT & 2009-02-12 & 54874.1 & 4.6 & 21.056 \\
00090007037 & WT & 2009-02-22 & 54884.5 & 4.6 & 31.431 \\
00090007038 & WT & 2009-03-04 & 54894.5 & 3.9 & 41.408 \\
00090007039 & WT & 2009-03-13 & 54904.0 & 4.1 & 50.912 \\
00090007040 & WT & 2009-03-24 & 54914.8 & 4.2 & 61.763 \\
00030956054 & WT & 2009-09-30 & 55104.5 & 3.3 & 251.428 \\

\hline
\label{ta:obs09}
\end{tabular}
\end{center}
\end{table}

\begin{table}[t]
\begin{center}
\caption{Burst statistics of magnetars}
\begin{tabular}{cccccccc}

\hline\hline
Magnetar    & $\overline{T_{90}}$ & $\overline{t_r}$  & $\overline{t_f}$ & $\overline{\Gamma}$ & $B^{a}$           & $P^{a}$ & Reference\\
            &  (ms)               &  (ms)             &   (ms)           &                     & ($10^{14}$ G) & (s) &   \\

\hline

SGR 1806$-$20 & 161.8 & - & - & - & 21 & 7.6 & \citet{gkw+01} \\
SGR 1900+14 & 93.4 & - & - & - & 7.3 & 8.0 & \citet{gkw+01} \\
1E 2259+586 & 99.31 & 2.43 & 13.21 & 1.35 & 0.59 & 7.0 & \citet{gkw04} \\
1E 1547$-$5408 & 305 & 39 & 66 & 0.17 & 2.2 & 2.1 & this work\\

\hline
\label{ta:mags}
\end{tabular}
\end{center}
$^a$From the McGill AXP/SGR catalogue at http://www.physics.mcgill.ca/$\sim$pulsar/magnetar/main.html
\end{table}

\clearpage

\end{document}